\documentclass[12pt]{article}
\usepackage{ulem}
\usepackage{times}
\usepackage{color}
\usepackage{graphicx}
\usepackage{calc}
\usepackage{epsfig}
\usepackage{float}
\usepackage{amsfonts}
\usepackage[section]{placeins}
\usepackage{amsmath}
\usepackage[makeroom]{cancel}
\usepackage{wrapfig}

\begin{document}

\Large
\begin{center}
{\bf The role of the observer in goal-directed behavior}
\end{center}

\normalsize

\begin{center}
{\sl In\'es Samengo}
\end{center}

\parskip 10pt
\baselineskip 0.25in

\vspace*{2cm}

\noindent
{\bf Abstract}

\noindent
In goal-directed behavior, a large number of possible initial states end up in the pursued goal. The accompanying information loss implies that goal-oriented behavior is in one-to-one correspondence with an open subsystem whose entropy decreases in time. Yet ultimately, the laws of physics are reversible, so entropy variations are necessarily a consequence of the way a system is described. In order to reconcile different levels of description, systems capable of yielding goal-directed behavior must transfer the information about initial conditions to other degrees of freedom outside the boundaries of the agent. To operate steadily, they must consume ordered degrees of freedom provided as input, and be dispensed of disordered outputs that act as wastes from the point of view of the aimed objective. Broadly speaking, hence, goal-oriented behavior requires metabolism, even if conducted by non-living agents. Here I argue that a physical system may or may not display goal-directed behavior depending on what exactly is defined as the agent. The borders of the agent must be carefully tailored so as to entail the appropriate information balance sheet. In this game, observers play the role of tailors: They design agents by setting the limits of the system of interest. Their computation may be iterated to produce a hierarchy of ever more complex agents, aiming at increasingly sophisticated goals, as observed in darwinian evolution. Brain-guided subjects perform this creative observation task naturally, implying that the observation of goal-oriented behavior is a goal-oriented behavior in itself. Minds evolved to cut out pieces of reality and endow them with intentionality, because ascribing intentionality is an efficient way of modeling the world, and making predictions. One most remarkable agent of whom we have indisputable evidence of its goal-pursuing attitude is the self. Notably, this agent is simultaneously the subject and the object of observation.

\newpage

A bunch of nucleic acids swim among many other organic compounds forming a cytoplasmatic soup, and somehow, manage to arrange themselves into precisely the sequence required for DNA replication. Carbon dioxide molecules steadily stick to one another materializing a solid tree trunk out of a tiny seed. Owls eat the young bats with poor navigation ability, thereby improving the eco-location proficiency of the species. The neurons in a dog's brain fire precisely in the required sequence to have the dog bury its bone, hiding it from other dogs. The wheels, breaks, and clutch of a self-driving car coordinate their actions in order to reach the parking area of a soccer field, no matter the initial location of the car, nor the traffic along the way. The limbs of the Argentine soccer players display a complex pattern of movements that carry the ball, through kicks and headers, at Messi's feet in front of the keeper... kick... {\bf goal}!

This essay is about goals. In all the above examples, a collection of basic elements, following local and apparently purpose-less laws, manage to steer the value of certain variables into some desired regime. The initial state is rather arbitrary, and yet, the agents manage to adaptively select, out of many possible actions, the maneuvers that are suited to conduct the system to the desired goal. Throughout these seemingly intelligent choices, order appears to raise from disorder. Scattered nucleotides become DNA. Air and dust become trees. Owl hunger becomes sophisticated eco-location organs. Neural activity becomes a buried bone. A car anywhere in the city becomes a car at a specific location. A football anywhere in the stadium becomes a football in the goal. How do the components of each system know what to do, and what not to do, in order to reach the goal? This is the question that will entertain us here.

\begin{wrapfigure}{L}{0.5\textwidth}
\centerline{\includegraphics[keepaspectratio=true, clip = true,
scale = 0.45, angle = 0]{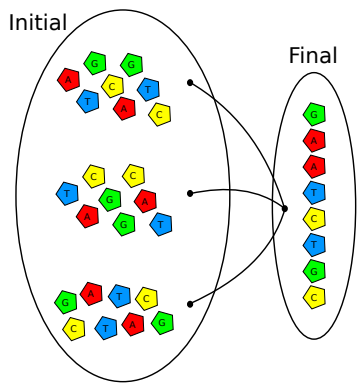}} \caption{\label{f1} In DNA replication, many initial states are mapped onto a single final state. The entropy is therefore high at the beginning and low at the end.}
\end{wrapfigure}
One important characteristic of goal-directed agents is that they are flexible: They reach the goal from multiple initial conditions, and are typically able to circumvent obstacles. For example, in DNA replication, the initial state is one out of many configurations in which nucleotides can be spatially distributed in a solution of organic compounds. The final state, the goal, is the precise spatial arrangement of those same nucleotides within the newly constructed DNA strand. In the soccer stadium, the ball may be initially in any location, the final state is the ball at the goal. Multiple initial states are hence mapped onto a single final state, as in Fig.~\ref{f1}. 
In physical terms, the non-injective nature of this mapping implies a reduction in entropy.

Admittedly, the final state need not be strictly unique. In DNA replication, permutations of equal nucleotides are still allowed in the final state, and occasionally, there might also be a few errors in the replication process. Dogs may consider more than a single location for the concealed bounty, and Messi may choose to shoot the ball anywhere inside the 24 ft wide by 8 ft high of the goal. Such restricted amounts of freedom, or even the occasional failures to reach the final state (shooting an own goal, for example), by no means compensate the abrupt reduction in entropy that takes place throughout the process.  In fact, were entropy not to decrease, the system would not exhibit goal-oriented behavior. 

We are used to associating entropy increments with information losses, and entropy reductions with information gains. Here I am taking the opposite view: Entropy reductions are associated with information losses. The two views are not incompatible, they simply refer to different things. The first case deals with a closed system and information about macroscopic variables. The second, with an open system and microscopic variables. When a closed macroscopic system evolves in the direction that maximizes the entropy of all compatible microscopic states (the usual case in closed thermodynamical systems), the final macro-state does not allow us to deduce the initial macro-state, since the mapping between them is non-injective. Were we to know the detailed final micro-state, however, we would be able to deduce the initial micro-state. When a goal-oriented system evolves in the direction of decreasing entropy, the final micro-state does not allow us to deduce the initial micro-state, but for a different reason: goal-oriented systems are open, and they interact with degrees of freedom we are not keeping track of. In this essay, the distinction between micro and macro-states is not emphasized, because the phenomena we deal with are not always divisible into separate scales.

The notion of goal-oriented behavior that is used here always brings about an entropy reduction. I now want to demonstrate the reciprocal statement: If a system reduces its entropy, a goal can be ascribed to the process. Therefore, entropy reduction and goal-oriented behavior are in a one-to-one correspondence.  The goal in question can always be defined by the restricted set of values that the variables acquire in the final state: the target DNA sequence, the buried bone, the ball at the goal. Of course, the reduction in entropy must first be verified: a broad set of initial states must evolve into a small final set. A car that in a single trial travels from one location to another is not guaranteed to be a self-driving car. Only if the initial location has proven to be arbitrary, and the traffic conditions variable, can goal-directed behavior be arrogated. 

The notion of entropy is subtle, since it not only characterizes a physical system, but also, the way it is described. When the universe is described at its utmost basic level (assuming there is one such level), all we have is a collection of fundamental particles evolving from some initial state. If the state of all particles is specified, the total entropy of the universe vanishes. Time reversibility of the laws of physics dictates entropy to remain zero for all past and future times. Therefore, there is no way to attain neither an increase nor a reduction in entropy. Energy dissipation and goal-directed behavior, hence, are absent from the complete description. We need to blur our point of view to give them a chance, either by restricting the description to macroscopic variables, or to subsystems. In fact, the main conclusion of this essay is that an observer with a very special point of view is required for agency to exist. 

If the information about the initial conditions is apparently lost in goal-oriented behavior, then such information must be somehow concealed in degrees of freedom we are not keeping track of.  They may have been moved into too microscopic states to be monitored, or into fluid degrees of freedom that, by the time the goal is reached, have already exited the subsystem under study. What we track, and what we ignore, hence, plays a crucial role in agency.

\begin{wrapfigure}{L}{0.5\textwidth}
\centerline{\includegraphics[keepaspectratio=true, clip = true,
scale = 0.25, angle = 0]{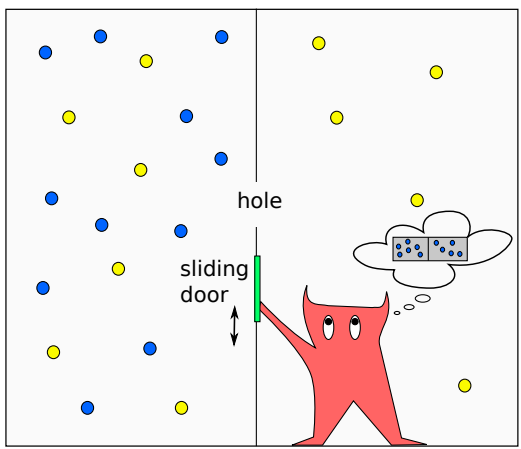}} \caption{\label{f1a} A demon controls the sliding door, allowing particles to pass from right to left, but not the other way round. The initial state of every molecule that the demon has already acted on (letting it pass or not) is recorded in its memory, and depicted in blue. }
\end{wrapfigure}
To be consistent with the second law of thermodynamics, processes where entropy decreases are only possible in open systems that somehow interact with the external world. Originally, they were supposed to require an energy influx. This is, however, not a necessary condition: Sometimes, the sole exchange of information suffices. A good example is Maxwell's Demon \cite{maxwell1908}. Suppose we have a gas enclosed in two adjacent chambers communicated by a small hole in the wall between them (Fig.~\ref{f1a}). The hole may or may not be covered by a sliding door controlled by a demon. Initially, both chambers have equal pressure and density. The demon then opens or closes the hole selectively, depending on whether a molecule approaches from one side, or the other. Molecules coming from the right are allowed to pass into the left chamber, but not the other way round. As time goes by, molecules accumulate on the left side, eventually leaving the right side empty. The collection of all gas molecules can be interpreted as performing goal-directed behavior: No matter the initial state, gas is gradually compressed into the left chamber. This final state can be conceived as a goal, and it comprises a reduction in entropy: initially each particle can be anywhere in the two chambers, and in the final state, they are all in the left side. Arrogating purpose, in this case, is to assume that the gas---who takes the role of {\sl the agent}---{\sl wants} to shrink. Other verbs may be used ({\sl tends to}, {\sl is inclined to}, etc.), but the phrasing is irrelevant. As uncanny as it may seem, arrogating purpose to the gas is a rather accurate description of the gas' phenomenology.

The gas $+$ demon is a toy model of a closed system, so no interaction with the outside world is allowed. To perform the task, the demon needs to acquire information on the location of each molecule approaching the hole, to then decide whether to let it pass or not. In a slightly modified version of this system, Bennett \cite{bennett1982} demonstrated that the storage of information in the demon's memory can be done with no energy expenditure, as long as the memory is initially blank, and there is plenty of storage capacity. The work required to move and stop the door, as well as the energy needed to measure the position of particles and to maintain the demon alive, can also be made as small as desired, simply diminishing mechanical friction, and moving slowly. The demon is however not allowed to delete the acquired information, because information erasure requires energy consumption, at a minimal cost of $k_{\rm B}T$ per erased bit \cite{landauer1961}. Therefore, as time goes by, the information of the initial location of each gas molecule is erased from the gas, and copied onto the demon's memory.

The gas gradually reduces its entropy only if we are careful to exclude the demon from what we define as {\sl the system}. If we include the demon (and its memory), entropy simply remains constant, since all the details of the initial state are still stored. Depending on the observer's choices, then, entropy may or may not decrease, meaning that arrogating agency may or may not be possible. 

A subsystem can only decrease its entropy if it somehow gets rid of initial conditions. In DNA replication, after the addition of each new nucleotide to the developing strand, the initial location of the free nucleotide determines the final configuration of the mediating enzimes, thereby transferring the information of the initial state to a change in the 3-dimensional configuration of nearby proteins. If enzimes are not restored to their functional state, the process cannot be iterated. So enzimes, in turn, must pass the information on somewhere else. This transfer is actually the important point in the emergence of goal-directed behavior. Energy consumption is only helpful if energy is {\sl degraded} in the process: ordered energy sources must be transformed into disordered products. In animal cells, order arrives as glucose and oxygen molecules. Disorder exits as carbon dioxide, water and faster molecular motion (heat). The input degrees of freedom, specifically in the case of glucose, are conformed of atoms tidily organized into large molecules. The output degrees of freedom  are transported by smaller molecules, amenable to be arranged in many more configurations. 

The laws of physics are ultimately reversible, so initial conditions cannot be truly erased, they can only be shuffled around. As an example, Edward Fredkin studied how non-dissipative systems, such as our universe, may perform the usual logical computations (AND, OR, etc), which are themselves non-invertible \cite{fredkin1982}. We know that 0 AND 1 = 0. However, knowing that the result of the operation is 0 does not suffice to identify the two input variables. If the computation is performed by an ultimately non-dissipative system, the information of the initial input variables must be somehow moved into some other variable, albeit perhaps not in a manner that is easily accessible. Fredkin's solution was to prove that  computing required some extra input variables, not needed for the computation per se, but mandatory for the information balance. When performing a single logical operation (say, for example, AND), the additional variables are in a well defined state (no uncertainty), and throughout the computation, they acquire so-called garbage values (garbage because they are not required to perform the computation), that represent those input degrees of freedom that cannot be deduced from the output. Copying part of the input into garbage variables ensures that no information is lost, and the computation becomes feasible in a non-dissipative physical substrate.

\begin{wrapfigure}{L}{0.5\textwidth}
\centerline{\includegraphics[keepaspectratio=true, clip = true,
scale = 0.4, angle = 0]{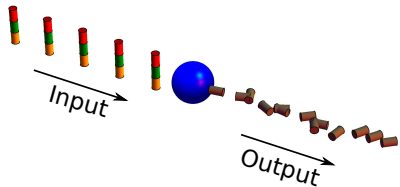}} \caption{\label{f2} Goal-directed systems (blue ball) eat up ordered degrees of freedom, and produce disordered degrees of freedom.}
\end{wrapfigure}
Ascribing agency is all about ignoring who really did the job (the Universe, to put it grandly), and arrogating intentionality to an entropy-reducing subsystem. The task of the observer is to design the borders of the subsystem so as to allow ordered degrees of freedom to be progressively incorporated, and/or disordered ones to be eliminated. If the goal is to be achieved repeatedly, a steady flow of order is required, as well as a regular garbage collection service. Purposeful agents, hence, only emerge from sub systems that eat up order (Fig.~\ref{f2}). Broadly speaking, they can be said to breathe, or to be endowed with metabolism,  even if they need not be alive in the biological sense.

Maxwell's demon hid the initial conditions of the gas in its memory. The dog, the self-driving car, and the soccer players, all hide their own initial state and that of the environment inside their memories. Memories can of course be erased, but erasures consume energy, and they are ultimately no more than flushing initial conditions into the high-entropy products of energy degradation. 
\begin{wrapfigure}{L}{0.5\textwidth}
\centerline{\includegraphics[keepaspectratio=true, clip = true,
scale = 0.8, angle = 0]{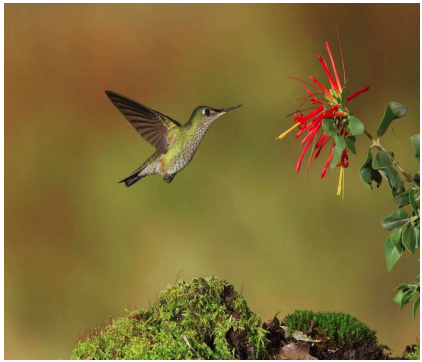}} \caption{\label{f3} Observers, just as photographers, selectively focus on an aspect of reality, to satisfy their cognitive appetite \cite{norretranders1999}. Potograph kindly supplied by Luc\'{\i}a Samengo.}
\end{wrapfigure}
For a long time, scientists failed to include memories as part of the systems under study, so goal-directed behavior sometimes appeared paradoxical. Here I argue that observers attribute agency by disregarding initial conditions. Of course, observers are free to delineate the borders of the subsystem under study as they wish. They can always shape the limits of what they define as {\sl the agent} in such a way as to have it do all sorts of wonderful things, as achieve goals, and reduce entropy. The agent must be fed with order, and the mess must be cleaned up, but still, it can be done. The natural question is therefore: What is interesting in goal-directed behavior if the observer is allowed to engineer the very definition of the agent, in order to get the desired result? Plants grow because what we define as a plant is the stuff that grows every spring, and not the dirt left on the ground every autumn. Species improve because we restrict the definition of a species to the material that a posteriori is seen as successful, and exclude the corpses left behind of those who failed. Cell division seems to be a productive business because the waste products are not defined to be part of cells. Returning to the question posed above ``How do the components of the system know what to do, and what not to do, in order to reach the goal?'', we can now provide an answer. Components know nothing, observers do. Just as photographers select an arbitrary plane in the visual world where to focus an image and engender a sharp object (Fig.~\ref{f3}), so do observers choose which variables compose the system, and which do not, so that a goal emerges.

Should we be amazed that the world we live in allows observers to create agents? Could we not live in a universe where assigning agency were downright impossible? I would be very much surprised if it were so. The impossibility to define goal-directed behavior would mean that no subsystems exist where entropy decreases. The global entropy growth that takes place in the whole universe should develop uniformly and monotonously all throughout space and time, allowing for no local oscillations. That would be ordered indeed! I do not expect disorder to arise in such an orderly manner. 

Within this picture, all the interesting events seem to take place in the observer's creative act. Any local decrement of entropy, no matter how trivial, appears to suffice for an observer to be able to ascribe agency. We are demanding little of the world, and a lot of the observer. But does the evolving world not have organizational merits of its own? If we look at the history of events taking place in our planet, as time goes by, agents seem to become increasingly sophisticated. Compare the strategy of a replicating DNA molecule in a bacterium with the one of Menelaus of Sparta to recover Helen of Troy, and thereby, ultimately manage bisexual reproduction. All the complexity of the bacterium's strategy is present in Menelaus', but not the other way round. Evolution seems to be striving towards what appears to be a runaway escalation of sophistication and design \cite{dennet1995}. Is the development of refined agents something that only depends on the observer's creativity, or is it something actually taking place independently of observers? 

\begin{wrapfigure}{L}{0.5\textwidth}
\centerline{\includegraphics[keepaspectratio=true, clip = true,
scale = 0.85, angle = 0]{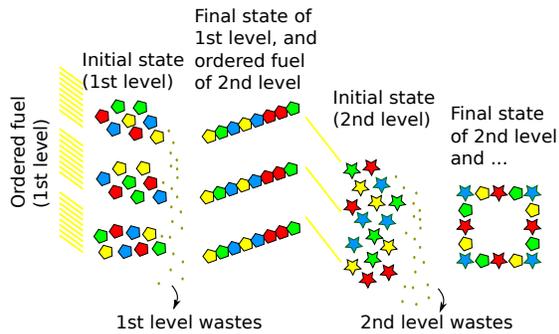}} \caption{\label{f4} The goal achieved by low-level agents can become the ordered fuel of higher-level agents.}
\end{wrapfigure}

Merits are shared, I believe. Observers produce agents. In the absence of agents, no subsystems are cut out of the wholeness of the cosmos, and complexity cannot be measured. Once observers are in play, even if they might have never intended it, it turns out that the computation they perform is liable to iteration. Agency allocation implies that many equal final states are produced from many different initial states. The final states are similar to one another, and similarity is a form of order. Such final states can therefore be used as the ordered degrees of freedom that a higher-level process may use as fuel. This fuel needs to enter into a noisy system for higher-level agents to emerge (Fig.~\ref{f4}). 
Noise is typically instantiated by a changing environment, often in combination with the occasional mistakes that may have happened in the lower-level process, as mutations in DNA replication. By iterating the algorithm, profuse RNA replication in free solutions can be observed to give rise to prokaryote cells, who in turn evolve into eukaryotes, from which multi-cellular organisms appear, all the way up to the ever growing branches of the tree of life. In the way, conscious humans, civilization, and artificial intelligence emerge. As well as a lot of garbage, as environmentalists wisely remind us. 

The tree of life develops as a continuous process, irrespectively of whether observers interpret it or not. Observers are required to dissect it into agents, and to evaluate their sophistication. As we climb the ladder of evolutionary design, the fuel degrees of freedom diminish in number, and increase in complexity, the latter defined as the number of bits required to describe their inner structure. If the resources at the bottom level are finite, the process cannot be iterated indefinitely, since eventually, too few and too complex degrees of freedom may not be identifiable as multiple instances of one single ordered pattern. 
 
Brain-guided observers are continuously ascribing agency. They do so because the role of a brain is to model the world around its carrier, so that effective survival strategies can be implemented. They have evolved to do so. Mental models must capture the regularities of the world, and discard the noise. Here, noise is defined as the degrees of freedom that are irrelevant to predicting those features of the environment that affect the observer's fitness. It would be a waste of resources, if not impossible, for us to represent in our brains all what happens in a dog's brain. Much more efficient is to ascribe agency, and conclude that the dog {\sl wants} to bury the bone. We cannot follow the evolution of all the bats that were eaten by owls, we therefore conclude that the predation of owls {\sl sharpens} the eco-location capacity of bats. We do not care for the details with which self-driving cars are programmed, we just think of them as goal-directed. We need an economic description, so we assign agency. 

Observers do not assign agency to all the entropy-reducing systems they meet. Purpose is only arrogated to subsystems for which there is no evident source of order, or for sources that are too costly to represent. The cost of a representation is judged in terms of its contribution to prediction accuracy. If we only look at the gas controlled by Maxwell's demon, ignoring the demon itself, we conclude that the gas wants to shrink. If, however, the demon takes weekends off, the purposeful model of the gas loses accuracy. A more sophisticated representation discerning between week days and weekends is needed. Assigning agency may or may not be a convenient strategy, depending on the trade-off between the economy of the representation and the prediction errors it induces. Arrogating agency in excess, for example by believing that all what happens is maneuvered by some obscure intentionality, yields a poor prediction strategy.

Observation is the result of development: Observers learn how to observe, and they do so within the framework of learning theory \cite{mackay2003}. They are first exposed to multiple examples of the process, that act as the training set. Before learning, the final state can only be predicted from the initial one if all degrees of freedom are tracked - a representation capacity that observers typically lack. Making the best use of their resources, observers explore the power set of the system (the set of all subsets of the system) and search for some entropy-reducing subset from which an agent and a goal can be defined. They then discard the superfluous degrees of freedom, thereby compressing information. Yet, if the subsystem does indeed reduce entropy, they are still able to make predictions. With successful predictions the world begins to makes sense, so ascribing agency is in a way equivalent to constructing knowledge. In fact, the construction of knowledge can be argued to be the essence of a mind.

In the last paragraphs, we have been observing observers. In doing so, we have placed ourselves one step above the hierarchical ladder of observation. We have concluded that purposeful agents do not exist per se, they are a mental construct of observers. This view may be easily accepted when regarding the agents (let us call them ``zombies'') around us, but is more problematic when it applies to ourselves. In the end, we experience our own purposes in a most irrefutable manner. In the present context, a self with purposes is no more than a compressed representation of an observer. Within a physicalist's point of view, the observer of the self cannot be situated in any other place than in the same brain where the self emerges. Whether the observer coincides with the self, whether it only partially overlaps with it, whether it contains it, or is contained by it, I do not dare to assert. I conjecture, however, that brains create selves following the same principles with which they ascribe agency to external factors, the difference being that the creation of the self involves a vastly larger number of degrees of freedom. Those degrees of freedom, moreover, are typically only accessible to the local subsystem. They include the mental processes of which we have conscious access, encompassing external sensory input, and the detailed state of our body. The latter has been proposed as the base for emotion, and the higher-level neural patterns triggered by such a state, the base of feeling  \cite{damasio1999}.

Within the self, the subject and the object of observation seem to coincide, forming a strange loop. Douglas Hofstadter \cite{hofstadter2007} has suggested that the circular nature of the mind observing itself is essential to the self. I am not sure, however, whether this recursive hypothesis constitutes an actual explanation of the self, or simply a way to bind the two loose ends together and worry no more. It could also be the case that what we perceive as a unitary self is in fact a whole collection of disperse mental processes, inside which multiple observers coexist, although separately unaccessible. In the end, consciousness has been equated with complex and indivisible information processing \cite{tononi2008}, so accessing subprocesses may not be possible. Dennet however very strongly argues that if such mental subprocesses can be considered multiple observers \cite{dennet1991}, there is no such thing as a hierarchy, and even less, an ultimate observer. 

I am afraid I am unable to provide a finished picture of agency when going all the way up to the self. I hope, however, to have built a sensible image of other less intimate agents. The main conclusion of this essay is that the interesting part of agency is the observer. Physics does not make sense, observers make sense of it. Life does not have a meaning, {\sl we} give it a meaning. Life may not even be fundamentally different from non-life, it may just be a collection of subsystems that appear to have goals. Goal-directed behavior does not exist if we do not define our variables in such a way as to bring goals into existence. Bringing goals into existence is a task that brains perform naturally, because they have evolved to model and predict the future. One fundamental agent that has emerged inside each one of us is the self. The mechanisms behind this process remain unclear, but its evolutionary utility is undisputed. Were we not able to produce meaning, we would not manage to distinguish ourselves as a special part of the cosmos. We would not have a sense of identity, a sense of self-preservation, nor the ability to think. The self is required  to enunciate even the most basic statements, all the way up from {\sl cogito ergo sum}.

\end{document}